\documentclass[pra,twocolumn,showpacs,amsmath,amssymb,showkeys]{revtex4}
\usepackage{graphicx}
\usepackage{dcolumn}
\usepackage{bm}
\makeatletter
    
    \newcommand{\Rmnum}[1]{\expandafter\@slowromancap\romannumeral #1@}
    \makeatother

\begin{document}
\title{Implementation of local and high-fidelity quantum conditional phase gates in a scalable two-dimensional ion trap}

   \author{Ping Zou}
   \affiliation{Laboratory of Quantum Information Technology, ICMP and SPTE, South China Normal
   University, Guangzhou 510006, China}
   \author{Jian Xu}
   \affiliation{Laboratory of Quantum Information Technology, ICMP and SPTE, South China Normal
   University, Guangzhou 510006, China}
  \author{Wei Song}
   \affiliation{Laboratory of Quantum Information Technology, ICMP and SPTE, South China Normal
   University, Guangzhou 510006, China}
   \author{Shi-Liang Zhu}
   \email{shilzhu@yahoo.com.cn} \affiliation{Laboratory of Quantum Information Technology, ICMP and SPTE, South China Normal
   University, Guangzhou 510006, China}

  \begin{abstract}
  We propose a scheme to implement high-fidelity conditional phase gates on  pair of trapped ions
  immersed in a two-dimensional Coulomb crystal, using
  interaction mediated by all axial modes without side-band addressing.
  We show through numerical calculations that only local modes can be excited to achieve entangling
  gates through shaping the laser beams, so that the complexity of the quantum gate does not increase
 with the size of the system. These results suggest a promising  approach for realization of large scale
 fault-tolerant quantum computation in two dimensional  traps architecture.

\end{abstract}

\maketitle

\section{Introduction}

  The trapped ion system is one of the most promising candidates for
quantum computation\cite{wineland2008}. In the system, each qubit
is formed by the long lived  internal atomic levels of each ion,
and laser beams have been used to manipulate quantum states of
ions\cite{Cirac95,Wineland}.
Different theoretical schemes have been proposed for quantum gate
operations with the aid of the ion's collective
motions\cite{Cirac95,Wineland,sm1999,ri2003,Duan2004,Zhu2006} and
many have been experimentally demonstrated with small  numbers of
qubits \cite{Leibfried2003,Haffer,Haljan,Leibfried2005}.

 To
make a quantum computer have practical value, scalability of proposed
architectures to  many qubits is of central importance. However,
scaling up to tens of qubits in ion-trapped system proves to be a
great challenge and many efforts have been done in this area. The
main difficulty of the scalability is that the sideband addressing,
which is believed to be necessary in the gate operations proposed in
most schemes, becomes impossible as the ion number increase. There
are two promising approaches in literature to overcome this scalar
difficulty. One is to divide the trapped-ion quantum computer into
operation region and memory region. Quantum gate operations are
always implemented at the operation region where only a few ions
form a small linear ion chains, while ions can be shuttled between
the operation and memory regions \cite{Kielpinsky}. The other
promising approach is to use appropriately shaped laser pulses to
realize  high-fidelity conditional quantum gates, where all phonon
modes are considered in the gate operations even in a very large
linear ion crystal\cite{zhueur}. This approach may be significantly
strengthened in an improved scheme \cite{gdlin} where the tightly
confined transverse modes\cite{Zhu2006} are used to mediate quantum
gate operations and an anharmonic trap potential is proposed to
arrange the ions in a uniformly-spaced linear crystal. In this
approach, the gate operation  has a local character so that the
complexity of the operations does not depend on the chain
size\cite{Duan2004}.

   To date most studies are limited to system of the
linear trap with ions
laser-cooled and confined in one dimensional crystals,
little
attention has been
  payed to the two-dimensional crystals\cite{d.porras,Taylor,Buluta}.
Today penning trap  can confine a large number of ions ($10^4-10^6$)
by a potential with approximate cylindrical symmetry \cite{Mitchell,
itano}.
In penning trap, axial confinement is realized by a static electric field, while
radial trap potential is induced by the rotations of the ions under an
axial magnetic field.
Under certain parameters the ions will be arrayed in a two-dimensional plane
with separation of order of tens of microns, such that they are
individually addressable by optical means.
 Although the rotation of the ion crystal may induce additional technical difficulty in
the addressing of the ions, quantum manipulation of such qubits is still possible. Particularly
 it has been proposed
that the axial phonon modes, which are independent on the phonon
modes in the planar direction since the anharmonic effects of
phonon modes are negligible,  can be used in the conditional gate
operations\cite{d.porras}. However, the phonon modes in a
two-dimensional crystal are much more complicated than that of a
linear ion crystal, and such complex mode spectrum makes it more
difficult to resolve individual modes for quantum gate operations.
Clearly the two-dimensional penning trap may provide an almost ideal
system for scalable quantum computation and quantum simulation if
the aforementioned difficulty of sideband addressing can be
overcome.
Therefore, a practical scheme to achieve the conditional gate
operations without sideband addressing, which has not been touched
before in such large scale two-dimensional traps, is highly
needed.

 In this paper,  we propose a feasible scheme to implement  high-fidelity conditional phase gates
 without sideband addressing in the scalable two-dimensional traps.
   We investigate in detail the configurations of the
 ions as well as the normal modes of the two-dimensional crystals.
 This scheme uses the ion's motion along the axial direction.  Although
  the effects of phonon modes in the planar direction can be
  neglected, the axial phonon mode spectrum is still too complicated to be resolved in the gate operations.
  We show that
 the local modes suggested in Ref.\cite{Duan2004,zhueur,gdlin} can also be applied to mediate
the interaction between ion and laser
 beams,  so that the complexity of the quantum gate does not increase
 with the size of the system. Compared to the scheme proposed in Ref\cite{d.porras},
we can implement conditional gate faster by optimally controlling the pulse shape. Since the nearest-neighbor entangling quantum gates are sufficient
for fault-tolerant quantum computing in two dimensional trap,  in
sharp contrast to one dimensional chain where the nearest-neighbor
entangling quantum gates are insufficient\cite{Raussendorf}, the generalization to
two dimensional trap proposed here will be very useful.

The paper is organized as follows: In Sec. \Rmnum{2} we study the
equilibrium positions of the ions trapped in the two dimensional
penning trap. In Sec. \Rmnum{3} we present the normal modes of the
trapped ions  and especially investigate the properties of the
axial modes in detail. In Sec. \Rmnum{4} we show how to implement
the conditional phase gates with high fidelity using the axial
modes. A summary of our results and a conclusion are presented in
Sec. \Rmnum{5}.

\section{Equilibrium positions of ions in a two-dimensional harmonic trap}

The system we have in mind is a two-dimensional trap
\cite{Mitchell,itano,Drewsen} with $N$
ions confined in a harmonic external potential. Including their
mutual Coulomb repulsion interaction the potential energy of ions
is given by
\begin{equation}
\label{Potential} V=\frac{1}{2}M\sum_{n=1}^N (\omega^2_x
x_n^2+\omega^2_y y_n^2+\omega^2_z z_n^2) +\sum_{n<l}^N
\frac{e^2}{4\pi\epsilon_0}\frac{1}{r_{nl}},
\end{equation}
where ${\bf r}_n\equiv (x_n,y_n,z_n)$ denotes the position of the
$n$th ion, $M$ is the mass of each ion, $e$ is the electron
charge, $\epsilon_0$ is the permittivity of free space and
$r_{nl}=\sqrt{(x_n-x_l)^2+(y_n-y_l)^2+(z_n-z_l)^2}$ is the
distance between the $n$-th and $l$-th ions. Here $\omega_\eta$
$(\eta=x,y,z)$ is the trap frequency which characterizes the
strength of the trapping potential in the $\eta$ direction. In the
paper we assume that $\omega_x=\omega_y=\omega_r$ and
$\beta=\omega_z/\omega_r$. The parameter $\beta$ is a measure of
the trap anisotropy and it will uniquely determine the ionic
crystal structure  for given $N$ \cite{Daniel}.

 Similarly to what they do in Ref\cite{James} for the 1D linear case,
to determine the equilibrium positions of ions, we may expand the
position of the $n$th ion around its equilibrium position ${\bf
r}_n^{(0)}$ as ${\bf r}_n(t)\approx {\bf r}_n^{(0)}+{\bf q}_n (t)$,
where ${\bf q}_n (t)$ is a small displacement. The equilibrium
positions are determined by the conditions
$$\left[\frac{\partial V}{\partial {\bf r}_n}\right]_{{\bf r}_n={\bf r}_n^{(0)}}=0,$$
and its dimensionless form ${\bf u}_m={\bf r}_m^{(0)}/\ell$
$(m=1,2,\dots,N)$ with $\ell=\sqrt[3]{e^2/4\pi\epsilon_0
M\omega^2_r}$ can be described by a set of $3N$ equations given as
\begin{equation}
\label{U_x}
u^x_m-\sum_{n=1,n\not=m}^{N}\frac{(u^x_m-u^x_n)}{\ell_{nm}^3}=0,
\end{equation}
\begin{equation}
\label{U_y}
u^y_m-\sum_{n=1,n\not=m}^{N}\frac{(u^y_m-u^y_n)}{\ell_{nm}^3}=0,
\end{equation}
\begin{equation}
\label{U_z}
u^z_m-\frac{1}{\beta^2}\sum_{n=1,n\not=m}^{N}\frac{(u^z_m-u^z_n)}{\ell_{nm}^3}=0,
\end{equation}
where
$\ell_{nm}=\sqrt{(u^x_n-u^x_m)^2+(u^y_n-u^y_m)^2+(u^z_n-u^z_m)^2}$
denotes the dimensionless distance between ions $n$ and $m$.

\begin{figure}[tbp]
\centering
\includegraphics[scale=0.4]{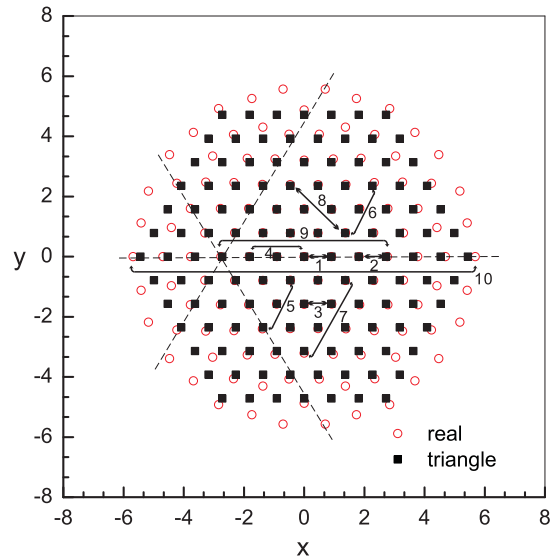}
\caption{ (Color online). The Configuration of ion crystals for the
number of ions $N=127.$ The circles represent the equilibrium
positions of ions obtained by numerical calculation. For comparison,
the solid squares denote the configuration of an ideal triangular
lattice. Conditional phase gates acting on 10 pairs of ions
connected by arrows are studied.} \label{Fig1}
\end{figure}

Determined by the number of ions and the trapped frequencies
$\omega_{x,y,z}$, the ions may form a string, a 2D or 3D ionic
crystals\cite{Daniel}. To individually address qubits by laser
beams, we usually assume ions are confined in a 1D or 2D crystals.
The 1D case has been extensively studied, but less work focussing
on the 2D traps\cite{d.porras,Buluta}. In this paper we pay
attention to the latter. Assume that $\omega_z$ is strong enough
such that the configuration of ions in a 2D plane is always stable
(the detailed requirement will be addressed below), then the
equilibrium positions are confined in $x-y$ plane and $u_m^z=0$
for an arbitrary ion. Under this condition, the equilibrium
positions in $x-y$ plane may be determined by solving
Eq.(\ref{U_x}) and Eq.(\ref{U_y}). The solutions are not unique,
for instance, ions for $N=4$ may locate at the vertexes of a
triangle and its center point or locate at the vertexes of a
square. However, both experiments and theories show that the
stable configuration of large number of ions forms a distortional
hexagonal lattice\cite{Daniel, Mitchell}, while the inner region
arranges in equilateral triangles. In Ref.\cite{Mitchell}
laser-cooled $^9$Be$^+$ ions in a two-dimensionally extended
lattice planes were directly observed, especially a single plane
hexagonal structure was observed, which is the configuration we
want to utilize. For concreteness, in this paper we study the case
where finite ions form the symmetric triangular lattice, such as the
ion numbers are $N=7,19,37,61,91,127,\cdots$. The configuration
for $N=127$ is shown in Fig. 1. In an ideal triangular lattice, the
equilibrium position ${\bf r}^{(0)}$ may be expressed as ${\bf r}
^{(0)}=[ (j+l/2)a_{0},\sqrt{3}a_0/2 ]$ with $a_0$ denoting the
lattice constant, $j$ and $l$ being integers. As expected the
equilibrium positions of ions obtained by numerical calculation
exhibit little deflection from the ideal configuration. As seen
from Fig. 1, the lattice spacings are not constant. As the ions
get farther from the center, the lattice spacings become larger;
however, it is almost constant near the crystal center and we get
a nice triangular lattice at the center that can be suitable for
the implementation of quantum computation.

\begin{figure}[tbp]
\centering
\includegraphics[scale=0.5,angle=0]{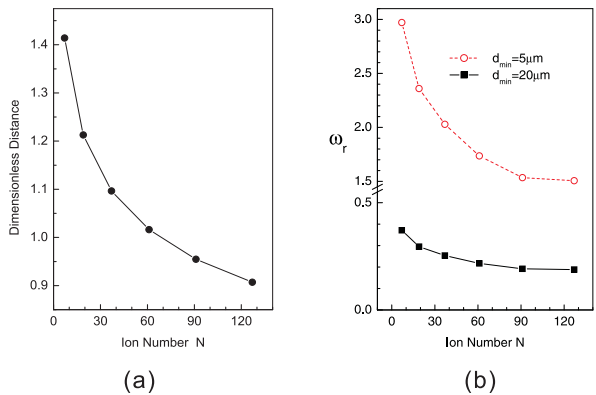}
\caption{ (a) Dimensionless minimum separation as a function of
total number of ions $N$ in a 2D harmonic  trap. (b) The trap
frequency $\omega_r/2\pi$ ($\mbox{MHz}$) required for the minimum of
distance $d_{min}=5\mu m$, $20\mu m$ as a function of $N$. Only the
data for $N=7,19,37,61,91,127$ are given. }
\end{figure}

 To manipulate quantum states of ions by laser beams, addressing individually
the ions by optical means are needed, thus
 the minimum interior spacing must be  larger
than the size of the focal spot of the laser beam which is about
$4\mu m$ in the experiments. The distances between the ions depend
on the strength of the confining potentials as well as the number
of ions. Actually the minimum value $d_{min}$ of the dimensionless
distance between two  adjacent ions descends as  the
total number of the ions is increased. By numerical fit from $N=7$ to $217$, we
find it obeys the relation $u_{min}\approx 1.995/N^{0.172}$, which
is a slower rate than  $2.018/N^{0.559} $\cite{James} in the
linear chain case. The typical scaled distances $u_{min}$ in 2D
trap with small ion number $N$ are shown in Fig. 2a.
 The relation between    inter
particle spacing $d_{min}=u_{min}\ell$ and the frequency
$\omega_r$ is $\omega_r=\sqrt{\frac{e^2 u_{min}^3}{4\pi
\varepsilon_0 M d_{min}^3}}$, so $\omega_r$ has an upper limit.
The examples with small $N$ are shown in Fig. 2b.

\section{Normal modes}

We now turn to investigate the normal modes that are useful in the
construction of the quantum operations in ion-trapped quantum
computation. The normal modes in a linear trap has been studied
in detail in Ref.\cite{James}, here we generalized the approach in one-dimensional crystal
to two-dimensional penning trap. In a realistic ion trap the ions will have some
nonzero temperature and will move around their equilibrium
positions. However, if the ions are sufficiently cold, we may
expand the potential $V$ under the harmonic approximation (i.e.,
the terms $O({\bf q}_n^3)$ can be neglected), and obtain $
V=1/2\sum_{n,m} {\bf q}_n {\bf q}_m[\partial^2 V/\partial {\bf
r}_n\partial {\bf r}_m]_0.$ Here the subscript $0$ denotes that
the double partial derivative is evaluated at ${\bf q}_n={\bf
q}_m=0$. In this case, the Lagrangian for the small oscillations
is given by

\begin{equation}
\label {L1} L=\frac{M}{2}\sum_{n=1}^N (\dot{{\bf
q}}_n)^2-\frac{1}{2}\sum_{n,m=1}^N {\bf q}_n{\bf q}_m\left[
\frac{\partial^2 V}{\partial {\bf r}_n {\bf r}_m} \right]_0.
\end{equation}
 The anharmonic
terms in $V$ will induce  small errors and can be neglected, which has been proved
in Ref. \cite{d.porras}. The double partial derivatives in
Eq.(\ref{L1}) may be calculated explicitly, then the Lagrangian
for the resulting small oscillations is found as

\begin{eqnarray*}
L= & & \frac{M}{2}\sum_{n=1}^N (\dot{{\bf
q}}_n)^2-\frac{M\omega_r^2}{2}\sum_{n,m=1}^N A_{nm}^{zz}q_n^z
q_m^z\\& & -\frac{M\omega_r^2}{2}\sum_{n,m=1}^N (A_{nm}^{xx} q_n^x
q_m^x+ 2A_{nm}^{x,y} q_n^x q_m^y +A_{nm}^{yy}q_n^y q_m^y),
\end{eqnarray*}

where
\begin{equation} A_{nm}^{xx}=\left\{
\begin{array}{ll} 1+\sum\limits^N_{p=1,p\not=
m}\frac{2(u^x_m-u^x_p)^2-(u^y_m-u^y_p)^2}{[(u^x_m-u^x_p)^2+(u^y_m-u^y_p)^2]^{5/2}},
& (n=m)
\\ \frac{-2(u^x_m-u^x_p)^2+(u^y_m-u^y_p)^2}{[(u^x_m-u^x_p)^2+(u^y_m-u^y_p)^2]^{5/2}}, & (n\not=m)
\end{array} \right.
\end{equation}

\begin{equation} A_{nm}^{xy}=\left\{
\begin{array}{ll} \sum\limits^N_{p=1,p\not=
m}\frac{3(u^x_m-u^x_p)(u^y_m-u^y_p)}{[(u^x_m-u^x_p)^2+(u^y_m-u^y_p)^2]^{5/2}},
& (n=m)
\\ \frac{-3(u^x_m-u^x_p)(u^y_m-u^y_p)}{[(u^x_m-u^x_p)^2+(u^y_m-u^y_p)^2]^{5/2}}, & (n\not=m)
\end{array} \right.
\end{equation}

\begin{equation} A_{nm}^{yy}=\left\{
\begin{array}{ll} 1+\sum\limits^N_{p=1,p\not=
m}\frac{-(u^x_m-u^x_p)^2+2(u^y_m-u^y_p)^2}{[(u^x_m-u^x_p)^2+(u^y_m-u^y_p)^2]^{5/2}},
& (n=m)
\\ \frac{(u^x_m-u^x_p)^2-2(u^y_m-u^y_p)^2}{[(u^x_m-u^x_p)^2+(u^y_m-u^y_p)^2]^{5/2}}, & (n\not=m)
\end{array} \right.
\end{equation}

 \begin{equation} A_{nm}^{zz}=\left\{
\begin{array}{ll} \beta^2-\sum\limits^N_{p=1,p\not=
m}\frac{1}{[(u^x_m-u^x_p)^2+(u^y_m-u^y_p)^2]^{3/2}}, & (n=m)
\\ \frac{1}{[(u^x_m-u^x_p)^2+(u^y_m-u^y_p)^2]^{3/2}}. & (n\not=m)
\end{array} \right.
\end{equation}

\begin{figure}[tbp]
\centering
\includegraphics[height=6cm,width=8cm]{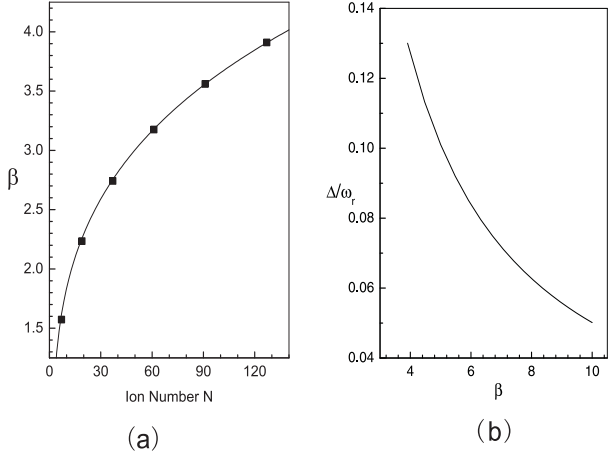}
\caption{ (a)The minimum $\beta$ required to confine all ions in a
$2D$ plane as a function of total number of ions in the Penning
trap. Only the data for $N=7,19,37,61,91,127$ are given. The line
represents the best-fit behavior of the trend. (b) The energy gap
of the axial modes between the center of mass mode and its nearest
mode as a function of $\beta$. }
\end{figure}
    In term of the normal phonon modes, the motional  Hamiltonian $H_0$
for $3N$ motional modes can be written as the standard form
$$
H_0=\sum_{k=1}^{3N} \hbar\omega_k (a_k^\dagger a_k+\frac{1}{2}),
$$
 which includes three possible polarization: axial ($\bf{z}$) modes  and two
 in-plane ($\bf{x},\bf{y}$) modes with   the eigen-frequency $\omega_k$.
 The axial  and in-plane modes are
 independent, and
 the axial modes are determined by the matrix
$A^{zz}$.  The eigen-frequencies
$\omega_{z,k}\equiv\sqrt{\mu_{z,k}}\omega_r$ and eigenvectors ${\bf
b}_j^{z,k}$ of the axial modes are obtained from diagonalization of
the matrix $A^{zz}$ with $\sum_n A^{zz}_{nj}{\bf
b}_n^{z,k}=\mu_{z,k} {\bf b}_j^{z,k}$. To make ions  stable in the
plane, all the angular frequency of the axial phonon modes
$\omega_{z,k}$, then the eigenvalues $\mu_{z,k}$ of the matrix
$A^{zz}(\beta)$ must be positive. Therefore, the equation
$\mu_{z,k}=0$ can be considered as the critical equation to
determine the stability of the two-dimensional configuration. After
solving this equation by numerical calculation, the critical values
of $\beta$ for different ionic number are shown in Fig. 3a. The
trend can be well described by a simple pow law
$\beta=\sqrt{1.073(N-2)^{0.55}}$, approximately coinciding with the
results that others obtained  when researching the structural
transition of the crystalline confined system, such as the empirical
scaling laws of Ref.\cite{schiffer}, and  the theoretic estimation
of Ref.\cite{Daniel}.
 So the critical value of $\beta$ is the
phase transition point that
 the 3D ionic structures become oblate to the $x-y$ plane, forming  a distorted 2D hexagonal
lattice. In the following we assume that the parameter $\beta$ is
larger then the critical values.

 The spectrum of the axial modes exhibits  some typical features: the center of mass mode at $\omega_z$ is the highest
frequency axial mode, while  the frequency splitting between the
center of mass mode and the second-to highest mode is the maximum,
and the splitting
 will become smaller when
increasing $\beta$,  as shown in Fig. 3b. We will impose the axial
modes  to implement quantum conditional phase gate in the next
section.



\section{Conditional phase flip gates}

\begin{figure}[tbp]
\centering
\includegraphics[scale=0.5,angle=0]{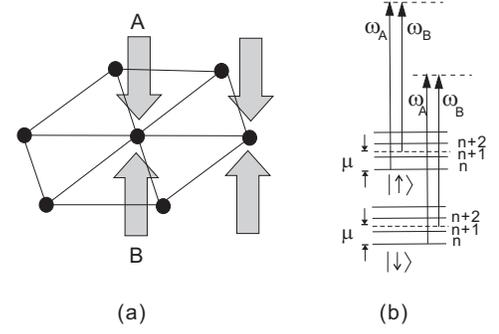}
 \caption{(a)Quantum gate in a penning trap:
electromagnetic fields induce a state-dependent dipole force on two
ions in a triangular lattice. (b) A $\sigma_z$-dependent force is driven
by electromagnetic fields with two frequencies separated by $\mu$,
These fields couple the qubit states to the excited states with different coupling strengths,
producing a differential AC Stark shift that oscillates at $\mu$. The two fields must have a non-zero
wavevector difference along the axial direction\cite{Lee}.}
\end{figure}
We now start to address quantum gate operations in this two
dimensional trap, mediated by many axial phonon modes.  To perform
quantum gates, we use the ac-Stark shift from two propagating
laser beams with a relative angle and detuning to apply
spin-dependent force on the ions along the $z$ direction(see FIG.4)
as they did
 in experiments \cite{Leibfried2003,Haljan,Lee}.  Under the Lamb-Dicke condition the Hamiltonian of the
system in the rotating-wave approximation
 becomes\cite{ri2003,zhueur}
\begin{equation}
\label{H_int} H =-\sum_{n,k=1}^N F_n(t) g_n^{k} (a_k^\dagger
e^{i\omega_{z,k} t} + a_k e^{-i\omega_{z,k} t})\sigma_n^z,
\end{equation}
where  $\sigma_n^z=|0\rangle\langle0|-|1\rangle\langle1|$ is the
Pauli operator, $F_n(t)$ is the spin-dependent force acting on the
$n$th ion, and $g_n^{k}=\sqrt{\hbar/2M\omega_{z,k}}{\bf
b}_n^{z,k}$ is the coupling constant between the $n$th ion and the
$k$th phonon mode, ${\bf b}_n^{z,k}$ are the eigenvectors of the
matrix $A^{zz}$ defined in the previous section. The
spin-dependent force has the form of $\Omega \sin(\mu t)$ deriving
from the ac Stark shift from the Raman laser beams, where $\mu$ is
the detuning between the Raman laser beams and $\Omega$ is the two
photon Rabi frequency. The evolution operator determined by
Hamiltonian (\ref{H_int}) can be explicitly found
as\cite{Zhu_PRL2003}
\begin{equation}
\label{U2} U(\tau) = e^{[ i\sum_n
\phi_n(\tau)\sigma_n^z+i\sum_{l<n}\phi_{ln}
(\tau)\sigma_l^z\sigma_n^z ]},
\end{equation}
where $\phi_n(\tau)=
\sum_k[\alpha_n^k(\tau)a_k^\dagger-\alpha_n^{k *}(\tau) a_k]$ with
 $\alpha_n^k(\tau)=\frac{i}{\hbar}\int_0^\tau F_n(t) g_n^{k}
e^{i\omega_{z,k} t} dt$ representing the residual entanglement
between ion $n$ and phonon mode $k$, and
$\phi_{ln}(\tau)=\frac{2}{\hbar^2}\int_0^\tau dt_2\int_0^{t_2}
\sum_{k} F_l(t_2)g_l^{k} g_n^{k} F_n (t_1) \sin\omega_{z,k}
(t_2-t_1) dt_1$ characterizes the effective qubit-qubit
interaction between ions $l$ and $n$. Note that the phase
$\phi_{ln}(\tau)$ is actually a kind of unconventional geometric
phase factor\cite{Zhu_PRL2003,Sjoqvist}, which has been
experimentally demonstrated to have advantage of
high-fidelity\cite{Leibfried2003,Du}.

\begin{table}[tbp]
  \begin{tabular}{|c|ccc|ccc|}
  \hline
  pair & \multicolumn{3}{c|}{$\omega_r/2\pi=0.2$} &
\multicolumn{3}{c|}{$\omega_r/2\pi=1.0$}\\
\cline{2-7}
   number & $F$ & $\mu$ & $\Omega^M$
&$F$ & $\mu$ &$\Omega^M$ \\
\hline
 1 & 0.999 & 10.033 & 0.177 & 0.994 &  10.076 & 0.443  \\
\hline
 2 &   0.999 & 10.033 & 0.178 & 0.989 & 10.076 & 0.442
\\
\hline
 3 & 0.999 & 10.033 & 0.173 & 0.990 & 10.045& 0.356
\\
\hline
 4 & 0.998 & 10.033  & 0.363  & 0.989 & 10.045 & 0.443
\\
\hline
 5 & 0.998 & 10.033  & 0.169  &  0.985& 10.076 & 0.577
\\
\hline
  6 & 0.998  & 10.033 & 0.369 & 0.973 & 10.025 & 0.534
\\
\hline
 7 & 0.997 &10.033  & 0.566 & 0.978 & 10.025 & 0.489
\\
\hline
 8 & 0.997 & 10.033  &  0.492 & 0.979  & 10.025 & 0.505
\\
\hline 9 & 0.997 &  10.087 & 2.768  & 0.940 & 10.076 & 1.090
\\
\hline
 10 & 0.995 & 10.040 & 4.040 & 0.690 & 10.076 & 1.315
\\
\hline
     \end{tabular}
  \caption{The gate fidelity $F$ for $10$ pairs of ions labeled in Fig.1  with  the same
$\omega_z$($10\mbox{MHz}$) but two different $\omega_r$. The units
for the detuning $\mu$ and the maximum Rabi frequency $\Omega^M$
are $1.0\mbox{MHz}$. }
     \end{table}

Eq.(\ref{U2}) is actually a conditional phase flip (CPF) gate acting
on ions $n$ and $l$ if $\phi_n(\tau)=\phi_l(\tau)=0$ and
$\phi_{nl}(\tau)=\pi/4$. To implement a CPF gate on arbitrary ions
$n$ and $l$, we make the spin dependent force  to be nonzero only
for these  two ions with $F_n(t)=F_l(t)=F(t)$, and set  the gate
time  $\tau$ so that $\phi_n(\tau)=\phi_l(\tau)=0$ and
$\phi_{nl}(\tau)=\pi/4$. In principle, it is always possible to
satisfy these constraints by designing a sufficiently complicated
pulse shape for the force.

Taking the method in Ref. \cite{zhueur}, we use a simple sequence of
laser pulses which take minimum experimental control. A
continuous-wave laser beam is chopped into  $m$ equal-time segments,
with a constant but controllable Rabi  frequency $\Omega_p$ for the
$p$th ($p=1,2,\cdots,m$) segment. The  spin dependent force $F(t)$
then takes the form $F(t)=\Omega_p\sin(\mu t)$ for the time interval
$(p-1)\tau/m\leq t < p\tau/m$. With  a small number of control
parameters  $\Omega_p$, $\alpha_n^k(\tau)$ (then $\phi_i(\tau)$) in
the evolution operation may not be  exact zero, but as long as they
are small enough, we still can   get a high-fidelity gate. The task
is to find an optimal  control laser beam parameters sequence
$\Omega_p$ and $\mu$ to  get a small enough infidelity $F_{in}$
($\equiv 1-F$ with $F$ being the gate fidelity). The gate fidelity
is defined as $F=\langle\Psi_f|\rho_r|\Psi_f\rangle$, where
$|\Psi_f\rangle$ is the ideal output state after implement the CPF
gate on the initial state $|\Psi_0\rangle$ which chosen typically as
$|\Psi_0\rangle=(|0\rangle_i+|1\rangle_i)\otimes(|0\rangle_j+|1\rangle_j)/2$,
and the density operator $\rho_r=Tr_m(U(\tau)(|\Psi_0\rangle\langle
\Psi_0|\otimes \rho^B)U(\tau)^\dagger)$ with tracing over the
motional states   of all the ions. Here we have assume that the
phonon modes are initially in thermal state $\rho^B$ with an
effective temperature $T$.

\begin{figure}[tbp]
\centering
\includegraphics[height=6cm,width=8cm]{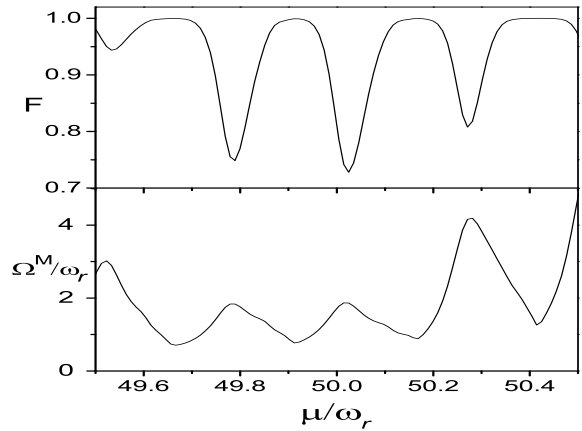}
 \caption{The gate fidelity $F$ and the maximum Rabi frequency $\Omega^M$ as a function of the
 laser detuning $\mu$ for ions pair $1$
with $\omega_r/2\pi=0.2\mbox{MHz}$, $\omega_z/2\pi=10.0\mbox{MHz}$
and gate time $50\mu s$.}
\end{figure}

Interestingly, we find that a very simple sequence of laser pulses
( such as $m=5$) can ensure a good gate fidelity even for the
two-dimensional trap. For concreteness, we calculate numerically
the CPF gates between $10$ pairs of ions labeled in Fig. 1, with
the same gate time $\tau=50\mu s$ and
$\omega_z/2\pi=10.0\mbox{MHz}$ but two different $\omega_r$. The
results are shown in Table  \Rmnum{1}. In Table \Rmnum{1}, we list
the optimal fidelity $F$, the corresponding detuning $\mu$ and the
maximum Rabi frequency $\Omega^M$ for each gate, where
$\Omega^M=$Max\{$\Omega_p$\} can be used to characterize the
maximum laser power required in the gate operations. In the
numerical calculation, we first calculate the fidelity and the
required Rabi frequency for a CPF gate as a function of the
detuning. As an example, the results for ion pair $1$ with
$\omega_r=0.2\mbox{MHz}$ is shown in Fig. 5, and for any other
gate operations the relation between the fidelity and the detuning
is similar, so fidelity has a local maximum near the detuning
$\mu=\omega_z$.  The data in Table \Rmnum{1} are derived from this
approach.

In Table \Rmnum{1}, the dimensionless distance between the pairs of ions is increasing in sequence. We observe several main features:
(i)  For both values of $\omega_r$, it is clear that
the gate fidelities decrease with the increasing of the distance
between the pair of ions. Especially for $\omega_r=1.0 MHz$, the
fidelity decreases relatively quickly with the distance between
ions, for example, the fidelity is about $0.994$ for the 1st pair of
ions, while it becomes $0.690$ for the 10th pair of ions. In
addition, the laser powers required for the CPF gates generally
increase with the distance between the pair of ions. (ii) To
understand the influence of  minimum ionic separation to the gate
fidelity, all gates are performed with two different trap frequency
$\omega_r$. One frequency is set as $\omega_r/2\pi=0.2\mbox{MHz}$,
and the other $\omega_r/2\pi=1.0\mbox{MHz}$, then the corresponding
minimum separation are $19.15\mu m$ and $6.57\mu m$, respectively.
From the table we can see that under condition of the same gate
time and $\omega_z$, gate realized in the trap with larger minimum
interior spacing will gain a higher fidelity. The physics behind it
will be addressed late. On the other hand, it is worth to pointing
out that, for pairs with lager distance, higher fidelity can be
gained with lower $\omega_r$ or longer gate time, for example,
taking gate time $\tau=200\mu s$, the fidelity for the ion pair $10$
with $\omega_r/2\pi=1.0\mbox{MHz}$ will be above $0.995$.

\begin{figure}[tbp]
\centering
\includegraphics[scale=0.6,angle=0]{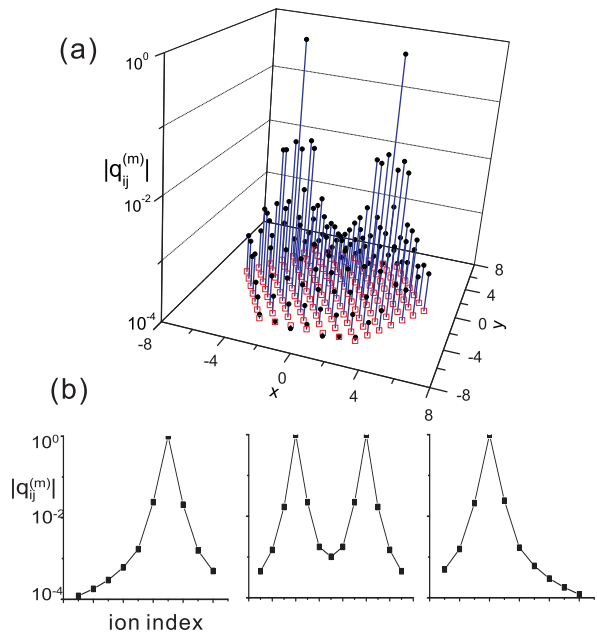}
 \caption{(Color online). The relative response of the ions for the gate operation.
(a) The points (blank squares) in the $x-y$ plane represent the
ions in the equilibrium positions and the
 $z$ values (filled circles) are the maximal deviation $q_{ij}^{(m)}$ from the equilibrium position of the ions during the
gate operation.
(b) The response of the ions located along the three lines drawn
in Fig. 1.  The $|q_{ij}^{(m)}|$ for the target ions have been
normalized to 1. The fast decay of the response as one moves away
from the target ions  shows that the gate operation involves
vibration of only local ions.}
\end{figure}

The physics behind such high-fidelity is that the gate has a local
character\cite{gdlin}, i.e., the  contribution to the CPF gate
comes primarily from the spin-dependent oscillations of the ions
that are close to the  target ions. To demonstrate the local
properties of the quantum gates, we plot in Fig. 6 the response of
each ion during the gate operation between the $9$th ion pair with
$\omega_r/2\pi=0.2\mbox{MHz}$. The figure explicitly exhibits the
ultra fast decay of the ions response from the target ions. So the
noises from the ionic vibration come just from the nearby ions. We
have shown in Table \Rmnum{1} that the fidelity would be higher
for a smaller $\omega_r$, since the ion spacings are less in the
larger $\omega_r$. We have calculated the response of all ions for
different $\omega_r$, and find that $q^m_{ij}$ of the nearby of
the target ions for $\omega_r=1.0MHz$ is larger of about $10\%$
than that of  $\omega_r=0.2 MHz$. So one can understand that the
fidelity would be lower by the increasing of the frequency
$\omega_r$. Attributing to this local character of the gate
operations, two notable advantages appear for the quantum
computation: (i) the complexity of a gate operation in the
two-dimensional traps does not depend on the system size, in sharp
contrast to the schemes based on side-band addressing. (ii) The gates
on the ions in different regions in the two-dimensional traps can
be performed in parallel.

The generalization of local gate operations from the linear
trap\cite{gdlin,Zhu2006} to two dimensional trap may have
advantages in the  fault-tolerant quantum computation. It is well
known that the set of quantum gates with local entangling quantum
gates and single-bit gates are universal for quantum computation.
Furthermore, it would be useful for large-scale fault-tolerant
quantum computing if the entangling quantum gates can be realized
for any pair of qubits. However, although the quantum gates for
any pair of ions in a large scale two dimensional trap are
possible, they are slow and costly in the sense that the laser
powers should be larger and the control parameters should be increased
if the distance between ions are large. In this case, an important
question is whether it is still possible to perform large-scale
fault-tolerant quantum computing with quantum gates that perform
only on the ions with the distance much smaller than the size of
the trap. The recent studies\cite{Raussendorf} have shown that in
a two-dimensional lattice, even with only nearest-neighbor quantum
gates, the error threshold can be still close to a percent level,
which is basically as good as the case with arbitrary nonlocal
gates, while the nearest-neighbor quantum gates are not sufficient
for fault-torrent quantum computing  in one dimensional chain. So
the generalization of trapped ion quantum computation to two
dimensional trap is highly deserved.

\section{Conclusions}
   In summary,  We  have explicitly exhibited the  configuration of the two dimensional trap and the
equilibrium positions of the ions trapped
 in small ion number by numerical simulation, and then examined the
condition on which all ions are confined in a plane.
 After investigating the properties of the axial modes of the trapped
ions,
 we have shown through explicit examples and
calculations that it is
  feasible to perform high fidelity CPF gate in the penning traps using the axial modes,
  and the gate  has a local character as in the linear chain
  case.
  The results suggest a realistic prospect for realization of
  large scale quantum computation in the penning traps architecture.

 We thank L. M. Duan, G. D. Lin, and M. Yang
for helpful discussions. The work was supported by  NSF of China
under Grant No. 10674049 and the State Key Program for Basic
Research of China (Nos.2006CB921801 and 2007CB925204).


\begin{thebibliography}{99}


\bibitem{wineland2008} For recent review of trapped ion quantum computation,
see R. Blatt and D. Wineland, Nature (London) \textbf{453}, 1008
(2008);  H. Haeffner, C.F. Roos, and R. Blatt,  Phys. Rep. {\bf
469}, 155-203 (2008).


\bibitem{Cirac95} J. I. Cirac and P. Zoller, Phys. Rev. Lett. {\bf
74}, 4091 (1995).

\bibitem{Wineland} D. Wineland, C. Monroe, W.M. Itano, D. Leibfried, B. King,
and D.M. Meekhof, J. Res. Natl. Inst. Stand. Technol. {\bf 103},
259 (1998).


\bibitem{sm1999} A. Sorensen and K. Molmer, Phys. Rev. Lett. {\bf 82},
1971 (1999); Phys. Rev. A {\bf 62}, 022311 (2000).

\bibitem{ri2003} J. J. Garcia-Ripoll, P. Zoller, and J. I. Cirac,  Phys.
Rev. Lett. {\bf 91}, 157901 (2003); Phys. Rev. A. {\bf 71}, 062309
(2005).

\bibitem{Duan2004} L.-M. Duan, Phys. Rev. Lett. {\bf  93}, 100502
(2004).

\bibitem{Zhu2006} S. L. Zhu, C. Monroe, and L. -M. Duan  Phys. Rev. Lett. {\bf  97},
050505 (2006).


\bibitem{Leibfried2003} D. Leibfried, B. DeMarco, V. Meyer, D. Lucas,
M. Barrett, J. Britton, W. M. Itano, B. Jelenkovic, C. Langer, T.
Rosenband, and D. J. Wineland, Nature (London) {\bf 422}, 412
(2003).

\bibitem{Du} J. Du, P. Zou, and Z. D. Wang, Phys. Rev. A {\bf 74}, 020302
(2006).


\bibitem{Haljan} P. C. Haljan,  K.-A. Brickman, L. Deslauriers, P. J. Lee, and
C. Monroe, Phys. Rev. Lett. {\bf 94}, 153602 (2005).

\bibitem{Haffer} H. H\"{a}ffner,  W. H\"{a}nsel, C. F. Roos, J. Benhelm1, D.
Chek-al-kar, M. Chwalla, T. K\"{o}rberl, U. D. Rapol, M. Riebe, P.
O. Schmidt, C. Becher, O. G\"{u}hne, W. D\"{u}r, and R. Blatt,
 Nature (London) {\bf 438}, 643 (2005).



\bibitem{Leibfried2005} D. Leibfried, E. Knill, S. Seidelin, J. Britton, R. B. Blakestad, J. Chiaverini, D. B. Hume, W. M. Itano, J. D. Jost, C. Langer, R. Ozeri, R. Reichle, and D. J. Wineland, Nature (London) {\bf 438},
639 (2005).




\bibitem{Kielpinsky}D. Kielpinsky, C. Monroe, and D. Winelannd,
Nature (London) {\bf 417}, 709 (2002)

\bibitem{zhueur} S. -L Zhu, C. Monroe, and L.-M. Duan,   Europhys. Lett {\bf73}, 485
(2006).







\bibitem{gdlin} G. D. Lin, S. L. Zhu, R. Islam, K. Kim, M. S. Chang, S. Korenblit, C. Monre, and L. M. Duan,
 Europhys. Lett. (in press).

\bibitem{d.porras} D. Porras and J. I. Cirac, Phys. Rev. Lett. {\bf
96}, 250501 (2006).

\bibitem{Taylor} J. M. Taylor and T. Calarco, Phys. Rev. A {\bf
78}, 062331 (2008)

\bibitem{Buluta} I. M. Buluta, M. Kitaoka, S. Georgescu, and S.
Hasegawa, Phys. Rev. A {\bf 77}, 062320 (2008).



\bibitem{Mitchell}T. B. Mitchell, J. J. Bollinger, D. H. E. Dubin, X. -P.
Huang, W. M. itano, and R. H. Baughman, Science {\bf 282}, 1290
(1998)

\bibitem{itano} W. M. Itano, J. J. Bollinger, J. N. Tan, B. Jelenkovic,
X. -P. Huang, and D. J. Wineland, Science {\bf 279}, 686 (1998)


\bibitem{Drewsen} M. Drewsen, C. Brodersen, L, Hornekaer, J. S.
Hangst, and J. P. Schiffer, Phys. Rev. Lett. {\bf 81}, 2878 (1998).

\bibitem{Daniel} D. H. E. Dubin, Phys. Rev. Lett. {\bf 71}, 2753
(1993).

\bibitem{Raussendorf}R. Raussendorf and J. Harrington, Phys. Rev. Lett. {\bf 98},
190504 (2007).

\bibitem{James} D. F. V. James, Appl. Phys. B {\bf 66}, 181
(1998).



\bibitem{schiffer} J. P. Schiffer, Phys. Rev. Lett. {\bf 70}, 818,
(1993).

\bibitem{Lee} P. J. Lee, K.A. Brickman, L. Deslauriers, P. C. Haljan, L.M.Duan, and
C. Monroe, J. Opt. B: Quantum Semiclass.  7, S371 (2005).


\bibitem{Zhu_PRL2003} S. L. Zhu and Z. D. Wang, Phys. Rev. Lett.
{\bf 91}, 187902 (2003).

\bibitem{Sjoqvist} For a review of goemetric quantum computation, see E. Sj\"oqvist, Physics {\bf 1}, 35 (2008).



\end{thebibliography}
\end{document}